\documentclass[sigconf]{acmart}

\usepackage{multirow}
\usepackage{xcolor}
\usepackage{tabularray}
\usepackage{hyperref}
\usepackage{subcaption}

\usepackage{framed}
\usepackage[most]{tcolorbox}
\newtcolorbox{bluebox}[1][]{
  colback=blue!5!white,
  colframe=blue!75!black,
  boxsep=2pt,      
  top=2pt,         
  bottom=2pt,      
  left=5pt,        
  right=5pt,       
  width=\linewidth 
}

\usepackage{blindtext}
\usepackage{balance}
\balance

\definecolor{main}{HTML}{5989cf}    
\definecolor{sub}{HTML}{cde4ff}     

\definecolor{firebrick}{HTML}{B22222}
\definecolor{yellowgreen}{HTML}{9ACD32}
\definecolor{royalblue}{HTML}{4169E1}

\begin{document}
\title[Quality Requirements for Code]{Quality Requirements for Code: On the Untapped Potential in Maintainability Specifications}

\author{Markus Borg}
\orcid{XXX}
\affiliation{%
  \institution{CodeScene and Lund University}
  \city{Malmö}
  \country{Sweden}
}
\email{markus.borg@codescene.com}

\renewcommand{\shortauthors}{Borg}

\begin{abstract}
Quality requirements are critical for successful software engineering, with maintainability being a key internal quality. Despite significant attention in software metrics research, maintainability has attracted surprisingly little focus in the Requirements Engineering (RE) community. This position paper proposes a synergistic approach, combining code-oriented research with RE expertise, to create meaningful industrial impact. We introduce six illustrative use cases and propose three future research directions. Preliminary findings indicate that the established QUPER model, designed for setting quality targets, does not adequately address the unique aspects of maintainability.
\end{abstract}

\copyrightyear{2024}
\acmYear{2024}
\setcopyright{acmlicensed}\acmConference[ICSE 2024]{46th International Conference on Software Engineering}{April 2024}{Lisbon, Portugal}
\acmPrice{15.00}
\acmDOI{10.1145/3593434.3593480}
\acmISBN{979-8-4007-0044-6/23/06}

\keywords{maintainability, quality requirements, source code quality}

\maketitle

\section{Introduction}
Quality requirements are critical for successful software products and services. They encapsulate non-functional aspects such as performance, reliability, and usability, which are vital in shaping the user experience and ensuring that solutions match their intended purposes. Attention to quality requirements can help software companies differentiate a product in a competitive market, emphasizing not just what the software does but how well it does it.

While researchers have argued about the importance of quality requirements for decades, they still present challenges for industry practice. In a recent systematic literature review, Olsson \textit{et al.} present several challenges~\cite{olsson_systematic_2022}. They conclude that the specification of quality requirements is under-developed in industry with informal treatment. There is no clear best practice and they do not appear to get the particular treatment they deserve, i.e., they often are treated like functional requirements despite their different nature. Involved challenges include quality measurement, requirements specification, and setting appropriate quality targets. 

Maintainability is a critical quality of the long-term success of a software organization. Its importance has been stressed since the early days of software engineering~\cite{dresner_maintenance_1964}. As users and customers cannot directly observe its benefits, maintainability is referred to as an internal quality. Exploratory studies on industry practice suggest that internal quality requirements receive little attention in the field~\cite{berntsson_svensson_investigation_2013}. Based on an interview study with 11 companies, Berntsson Svensson \textit{et al.} conclude that ``\textit{quality requirements that are not visible to the customer, such as maintainability and testability, are more often dismissed}'' in business-to-business contexts~\cite{berntsson_svensson_quality_2011}.

Academic software maintainability research has largely focused on source code metrics. The foundational metric McCabe's cyclomatic complexity~\cite{mccabe_complexity_1976} --- proposed almost 50 years ago --- is Software Engineering (SE) common knowledge. Secondary~\cite{nunez-varela_source_2017,ardito_tool-based_2020}, and even tertiary~\cite{abbad-andaloussi_relationship_2023}, studies list numerous alternatives such as object-oriented and aspect-oriented metrics. Some researchers developed compound metrics, such as Oman and Hagemeister's maintainability index~\cite{oman_construction_1994}, the SQALE maintainability index~\cite{letouzey_sqale_2012} and CodeScene's Code Health~\cite{borg_requirements_2023}.

How to best use maintainability metrics in requirements specifications remains an open question. While many researchers studied metrics, the Requirements Engineering (RE) research community has not sufficiently investigated how to apply them. The new standard ISO/IEC~5055 Automated Source Code Analysis contains vague suggestions such as using ``shall not'' requirements and violation densities~\cite{international_organization_for_standardization_information_2021} --- research is needed to devise better guidance.

As highlighted by this novel MO2RE workshop, RE is an underappreciated topic in the broader SE community. We know that studies presented at ICSE are often closely connected to the source code level. To close the gap, we posit that research on code-oriented maintainability requirements could act as a bridge builder. Moreover, we argue that improving maintainability requirements has a large potential to support a wide range of situations in industry. To initiate the discussion, this position paper presents six illustrative use cases and outlines three directions for future research.

\section{Illustrative Use Cases} \label{sec:uc}
This section showcases six use cases that would benefit from explicit maintainability requirements. Each use case is introduced through a fictive example in a blue callout box.

\subsection{UC1: Navigating Complex Automotive Supply Chains}
\begin{bluebox}
Volvo Cars works with around a hundred tier-1 suppliers, each providing crucial software-intensive subsystems. Prioritizing the need for quick adaptability to market changes --- necessitating software evolution --- the company wants to ensure that all suppliers maintain high-quality source code.
\end{bluebox}

In the automotive industry, where software is now driving innovations, the tier-layered structure often leads to integration challenges. The Taylorism driving the tier-layered structure served well for mechanical components~\cite{rawlinson_taylorism_1996}, but the situation changed for evolving software-intensive systems --- as made evident in comparison to the high velocity demonstrated by the software-first company Tesla. In a traditional setting, with a multitude of suppliers, software with varying levels of internal quality will be integrated into a vehicle. Adding new features that involve several suppliers turns slow. Consequently, there is a trend among Original Equipment Manufacturers (OEM) to insource software development, thereby gaining greater control over software assets. Software sometimes becomes too important to let others do it.

\textbf{What if the OEMs could specify requirements for high code quality among their suppliers?} The suppliers could then self-assess their maintainability by using software tools to extract data from their code repositories. This evidence could be compiled in assurance cases to demonstrate that the maintainability requirements are met. OEMs could then trust that their suppliers can swiftly adapt to new business requirements and feature requests. Such requirements could also support supplier selection as OEMs could establish supplier relations --- or even partnerships --- based on the long-term maintainability of candidate suppliers. Volvo Cars would appreciate this\ldots

\subsection{UC2: Public Software Procurement}
\begin{bluebox}
The Swedish Forestry Agency is procuring a new automated reforestation planning tool to assist small-scale foresters in optimizing their efforts. In the wake of forestry digitalization, they know that major new features will evolve in the coming decade. The procured system must be easily extendable.
\end{bluebox}

Software procurement in the public sector is often fraught with challenges, necessitating robust RE expertise~\cite{borg_digitalization_2018}. In Sweden, several software projects have made national news headlines, e.g., Stockholm's School Platform, and major IT systems developed for the Swedish Police, the Transport Agency, the Board of Agriculture, and the Public Employment Service --- project overruns and problems post-launch have been common. Procurement is not just about listing desired features, but it must also ensure the longevity and sustainability of the system. In the EU, the multi-criteria method \textit{Most Economically Advantageous Tender} (MEAT)~\cite{lehtonen_choosing_2022} can be used for selection instead of just looking at price tags. 

\textbf{What if government agencies could specify maintainability requirements?} They could then go beyond just specifying features and also set clear expectations for long-term maintainability. Together with other system qualities such as security and performance, maintainability could be input to the MEAT method. Utilizing maintainability metrics, they could effectively guide the agencies' procurement decisions. Moreover, continuous monitoring of these metrics post-launch could provide a window into the black box of the supplier's software, ensuring adherence to the requirements. Failure to meet these requirements should be treated as a contract violation, akin to breaching a service-level agreement~\cite{trienekens_specification_2004}, thereby providing the agency some protection against future technical debt. The Swedish Forestry Agency would happily avoid ending up with yet another IT system failure\ldots

\subsection{UC3: Open-Source Software Selection}
\begin{bluebox}
Anna, a software architect at a medium-sized telecommunications company, is about to make an important decision that will significantly influence her product's software bill of materials for years to come. She must decide whether to include Google's Protobuf or Apache Avro for data serialization.
\end{bluebox}

Open-Source Software (OSS) is a fundamental component in modern software systems. Organizations can leverage community-driven solutions rather than developing all code in-house. However, selecting which OSS solutions to include in a system can be a pivotal decision. Choices range from widely-adopted libraries backed by major tech corporations like Alphabet or Meta, to specialized libraries from smaller niche communities. While the decision-making process will never be easy, there is initial academic work on supporting OSS selection, including actor-based, software-based, and orchestration-based health metrics~\cite{linaker_how_2022}.

\textbf{What if OSS projects had explicit maintainability requirements?} Introducing such standards would allow OSS projects to demonstrate their commitment to this crucial quality aspect --- which would help potential integrators prioritize more effectively. There are already GitHub status badges\footnote{For example, \url{https://github.com/chetanraj/awesome-github-badges}} showcasing metrics such as test coverage and code-smell densities, but expressing maintainability requirements using validated metrics could take the badges to a whole new level. Anna would have loved seeing such badges\ldots

\subsection{UC4: Consultancy Agreements for Legacy Software}
\begin{bluebox}
Accenture is in negotiations for a contract to maintain a legacy system for a large retail company. The potential contract is attractive, but what amount of technical debt is hidden in that old VB6 codebase? Accurately pricing the tender is difficult in the face of uncertainty.
\end{bluebox}

Several IT consultancy firms specialize in taking on long-term maintenance contracts for various systems. These contracts typically entail ensuring the continuous functioning of integrations and the regular release of security updates. However, accurately pricing such tenders is difficult due to the uncertainties inherent in poorly documented legacy systems. Consequently, consultancy firms must rely on expert judgment for their cost estimations. This process involves cost estimation by analogy based on past projects, considering factors such as the technology stack, codebase size, and previous maintenance history. There are severe risks of either a) underestimating costs, leading to narrow profit margins, or b) overestimating them, resulting in tenders that are prohibitively high and thus rejected.

\textbf{What if the legacy system had maintainability requirements?} Even if no assurance case was available, the consultancy firm could use a tool-based analysis to get an understanding of the status. They could then leverage this information to make more informed bids, aligning their maintenance strategies with evidence about the system's code quality. This would enable substantially more informed pricing negotiations. The consultancy would reduce the risk of taking on maintenance projects with unknown variables, whereas the client would obtain cost-effective maintenance of their system. A win-win situation for both Accenture and their clients\ldots

\subsection{UC5: Due Diligence Before Acquisition}
\begin{bluebox}
In the boardroom of Salesforce, executives are deliberating over the acquisition of a provider of conversational AI for customer service automation. The external quality of the solution is certainly impressive. But what about its internal quality?
\end{bluebox}

In the dynamic landscape of company acquisitions, due diligence is a critical process that supports informed decision-making. Traditionally, this process has focused on evaluating financial health, market position, and legal compliance. However, as software strongly drives innovation nowadays~\cite{andersson_software_2021}, the maintainability of companies' source code assets should receive a top priority. Acquiring a company whose code is easy to maintain and evolve would be much more favorable --- thus a tool-based code analysis should be an obvious complement to the traditional activities.

\textbf{What if due diligence could include verification of maintainability requirements?} If the company under scrutiny already had corresponding requirements and an accompanying assurance case, the code-oriented due diligence would be streamlined. Lacking such requirements, the potential acquirer --- or a third-party assessor conducting the due diligence --- could still specify maintainability expectations that the target company could relate to. Using tool-based measurements, objective metrics could be collected and compared to other pieces of software already in the buying company. The SalesForce CTO would happily have received such a report before entering the boardroom\ldots 

\subsection{UC6: Evergreen VCs in the Dragon's Den}
\begin{bluebox}
Behrooz, a partner at an evergreen VC firm, reflects on his recent meetings with startups. One of them caught his attention with their innovative approach. However, how mature are their engineering skills? Could he trust that they are building a sustainable software foundation from the start?
\end{bluebox}

The startup ecosystem, flourishing in today's economy, relies heavily on Venture Capitalists (VC) for growth and development. At the same time, startups and entrepreneurship are vital for the economic expansion of society --- and software startups are a big piece of this~\cite{klotins_software_2019}. The investment process typically progresses through stages, initially focusing on innovation and concept validation, and gradually shifting towards the scalability and sustainability of the business model. Evergreen funding models~\cite{halt_sources_2017}, characterized by a continuous injection of capital into businesses, diverge from traditional VC funds, which usually operate within a specific timeframe for returns. These evergreen models aim for sustainable and steady growth over rapid short-term gains. Consequently, VC firms utilizing evergreen funding must pay close attention to the maintainability of software startups' codebases.

\textbf{What if VC firms could specify maintainability requirements for startups and scaleups?} Evergreen VC firms could then set clear expectations for the maintainability of innovative software systems from the outset. If the startups already have such requirements, complemented with assurance cases, they could even present them as they pitch their ideas. While such detailed discussions might not be shared in the very first pitch, they certainly deserve to be mentioned in the negotiations that follow for promising entrepreneurs. Behrooz would appreciate hearing about maintainability when scrutinizing potential startups in his dragon's den\ldots

\section{Ongoing Work and Concluding Remarks}
Our research mission is to \textit{connect technical source code metrics with strategic decision-making in the executives' boardrooms}. We see RE as an appropriate vehicle for bridging these two worlds and creating a mutual understanding. RE is arguably the most multifaceted SE topic, through which different stakeholders can meet and discuss all sorts of engineering discussions. On the one hand, executives must learn that maintainable source code transcends mere vanity metrics for developers. On the other hand, developers must learn to drive metrics-based refactoring efforts according to the business strategy --- code that changes is code that matters.

We outline three important directions to evolve the use of maintainability requirements in industry practice. First, additional empirical research is needed to collect more evidence on the business value of high-quality source code. In our previous work, we have validated that the aggregated maintainability metric Code Health is correlated with fewer defects and faster issue resolution times in 39~proprietary projects~\cite{tornhill_code_2022}. Code Health is a numeric value between 1 (extremely poor quality) and 10 (top-notch quality). This metric integrates elements from ISO/IEC 5055's maintainability category~\cite{international_organization_for_standardization_information_2021} and adds design-level code smells such as God Class/Method, and Duplicated Code. We plan to extend our analysis with more projects and causal analysis of the observational data. 

\begin{figure}
    \centering
    \includegraphics[width=0.45\textwidth]{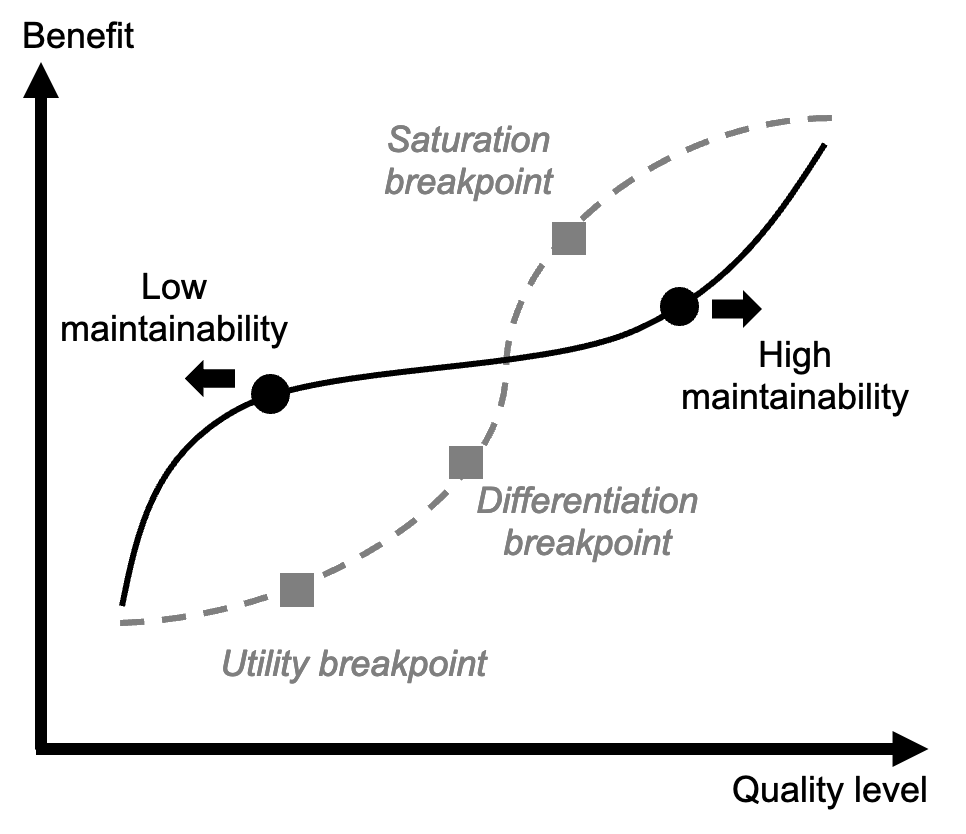}
    \caption{The benefit of maintainability vs. the QUPER model.}
    \label{fig:quper}
\end{figure}

Second, there is a need to model the value of maintainability based on the empirical data. Our ongoing work challenges the established QUPER model's \textit{benefit view} for maintainability, suggesting an alternative pattern in mapping quality to value. The QUPER model posits that the perceived benefit of quality improvements is a sigmoid function with three breakpoints~\cite{regnell_supporting_2008}, i.e., utility, differentiation, and saturation (see the gray dashed line in Fig.~\ref{fig:quper}). Our initial results in mapping file-level Code Health to value reveal a different story.

We recently proposed a model that approximates changes in defect density and issue resolution time as proportional changes in value creation~\cite{borg_increasing_2024}. Assuming that fewer defects and quicker issue resolution translate into higher value creation, our regression analysis explored these factors' variations across different code quality intervals. Our findings reveal significant non-linearities, especially at the quality spectrum's extremes, depicted as a logit function-like curve in Fig.~\ref{fig:quper}.

Our research offers a new perspective on maintainability, contrasting with the middle-range focus of the QUPER model. We hypothesize that code-level maintainability differences are the most important at the extreme ends of the quality scale. This aligns with the Broken Windows theory~\cite{hunt_pragmatic_1999}, which suggests that preventing the initial introduction of code smells is key to maintaining pristine code quality. On the other side of the quality scale, a poor-quality file may perpetuate further neglect and a costly quality decline. Further research is needed to substantiate these ideas. 

Third, there is a need to evolve requirements specifications for maintainability. Only very few studies have investigated how maintainability requirements are handled in industry practice~\cite{berntsson_svensson_quality_2011,berntsson_svensson_investigation_2013}. While research on code metrics is mature, there is a major need for industry-academia collaborations targeting how to best use the metrics in requirements specifications. While most SE practices must be tailored for specific contexts, we hypothesize that a general approach could be found to accommodate diverse needs as illustrated in Sec.~\ref{sec:uc}. The actual quality targets, on the other hand, must inevitably be selected on a case-by-case basis. 

Mainstream SE researchers will not suddenly come to RE venues. RE researchers must seek allies in the major SE venues! Evolving RE for maintainability is a task where we need to invite code-level researchers to co-learn for real impact. As so many times before, RE can act as the glue between different perspectives.

\balance
\bibliographystyle{ACM-Reference-Format}
\bibliography{requirements}

\end{document}